\documentclass[10pt,a4paper,showpacs,english,aps,prl,twocolumn]{revtex4}
\usepackage{amsmath}
\usepackage{graphicx}

\makeatletter

\begin{document}

\title{Pairing-based cooling of Fermi gases}

\author{M.J. Leskinen$^1$, J. Kinnunen$^{1,2}$, P. T\"{o}rm\"{a}$^1$}
\email{paivi.torma@phys.jyu.fi}
\affiliation{$^1$Department of Physics, Nanoscience Center, P.O.Box 35, 40014 University of Jyv\"{a}skyl\"{a}, Finland
\\$^2$ JILA and Department of Physics, University of Colorado at Boulder, CO 80309-0440, USA}


\begin{abstract}

We propose a pairing-based method for cooling an atomic Fermi gas. 
A three component (labels 1, 2, 3) mixture of Fermions is considered where the components 1 and 2 interact and, 
for instance, form 
pairs whereas the component 3 is in the normal state. For cooling, the components 2 and 3 are coupled 
by an electromagnetic field. 
Since the quasiparticle distributions in the paired and in the normal states are different, 
the coupling leads to cooling of the normal state even when initially $T_{paired}\geq T_{normal}$. 
The cooling efficiency is given by the pairing energy and by the linewidth of the coupling field. 
No superfluidity is required. The method has a conceptual analogy to cooling based on superconductor -- normal 
metal (SN) tunneling junctions. Main differences arise from the exact momentum conservation in the case of the 
field-matter coupling vs. non-conservation of momentum in the solid state tunneling process. Moreover, 
the role of processes that relax the energy conservation requirement in the tunneling, 
e.g. thermal fluctuations of an external reservoir, is 
now played by the linewidth of the field. The proposed method should be experimentally feasible 
due to its close connection to RF-spectroscopy of ultracold gases which is already in use.  
\end{abstract}

\maketitle

\begin{figure}
\label{schematic_cooling}
\centering
\includegraphics[width = 0.40\textwidth]{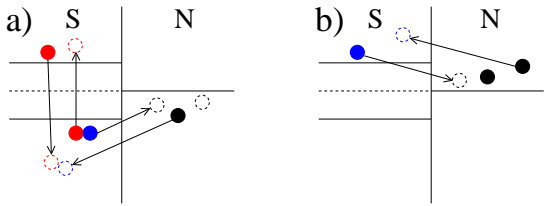}
\caption{(Color online) The normal state can be cooled by filling thermal holes (a) or 
removing hot particles (b). Net cooling is achieved because the thermal quasiparticle distributions
are different in the (N) normal and (S) paired (energy (pseudo)gap $\Delta$) states. 
The picture is schematic: there is no spatial tunneling barrier, instead, the normal and paired states
are coupled by an electromagnetic field.}
\end{figure}

Cooling is often a prerequisite for observing correlated quantum states of matter, 
as has been demonstrated, e.g., by the history of 
superconductivity, superfluidity and the quantum Hall effect. 
In the case of ultracold atoms, laser cooling and 
evaporative cooling have enabled the observation of Bose-Einstein condensates \cite{Anderson1995a, Davis1995a}, 
Mott insulator states \cite{Greiner2002a} and 
condensates of strongly interacting Fermi gases \cite{Jochim2003b, Greiner2003a, Regal2004a, Zwierlein2004a, 
Bartenstein2004a, Kinast2004a, Chin2004a, Zwierlein2005a}. The research is rapidly expanding to new directions such as
search for novel quantum states in optical lattices, p-wave pairing, and studies of 
multi-component mixtures of Fermions. To reach the required temperatures also in these new systems, 
cooling that goes beyond the presently available methods may become necessary. We propose a method for cooling 
fermionic atoms in a normal state by an internal tunneling contact to a paired state. 
The coexisting paired and normal states
can be realized by trapping a three-component Fermi gas and tuning the various inter-component couplings by 
Feshbach resonances.
The effective tunneling contact is realized by coupling the internal states of the atoms
by electromagnetic fields. The cooling method is closely related 
to RF-spectroscopy of the pairing gap in ultracold Fermi gases \cite{Chin2004a,Kinnunen2004b} and has an analogy to on-chip cooling 
methods of solid state structures via superconductor -- normal metal (SN) tunneling junctions 
\cite{Giazotto2006a}.

The basic principle of the proposed cooling is illustrated in Fig. 1. 
A three-component Fermi gas is confined to a certain spatial volume, e.g. by an (optical) trap or lattice. 
Components 1 and 2 interact, for instance via a Feshbach resonance, whereas the component 3 does 
not interact with the other components. 
Components 2 and 3 are coupled by an electromagnetic field, with a certain (effective) 
detuning $\delta=\omega_{field}-\omega_{23}$ and Rabi frequency $\Omega$. 
All atoms coexist in the same spatial volume. Conceptually, however, one 
can imagine a paired state (1 and 2) -- normal state (3) interface where the electromagnetic field creates a tunneling
coupling, as is displayed in Figs.\ 1a and 1b to illustrate different processes that contribute to the cooling. In Fig.1a, 
the detuning is chosen to be positive, $\delta>0$, which means that the field gives to the system excess energy 
which is sufficient for breaking a pair: one 
component of the pair (1) becomes an excitation in the paired gas whereas the other one (2) is transferred to the state 3
and fills a thermal hole in the normal
gas. The inverse process, which also occurs for the same detuning, is that an atom from the normal state joins a 
thermal quasiparticle in the paired state and forms a pair. However, due to the pairing (pseudo)gap, there are fewer 
quasiparticles in the paired state than in the normal state, see \cite{Mahan} and discussion after Eq.(\ref{DeltaN_2}), 
and this causes an imbalance between
the two processes and thus a net cooling effect. 
In Fig.1b, a thermal quasiparticle in the paired state is transferred to 
the normal state. To enable this, the field has to take the excess energy of the quasiparticle, therefore the process
occurs for $\delta<0$ (in contrast to Fig.\ 1a where $\delta>0$). Now the inverse process is 
the transfer of a hot quasiparticle away from the normal state. Again, the imbalance in the quasiparticle 
distributions leads to the desired cooling of the normal state. Note that the process corresponds to heat 
transfer between the paired and normal components, and no entropy is removed from the total system as will be 
discussed in detail below. In the following, we demonstrate the efficiency of the method, 
determine by numerical and analytical calculations the optimal parameters and
limitations for cooling, and discuss the connection and differences to SN tunneling junction 
coolers.

We consider a three-component gas, i.e.\ fermionic atoms in three different internal states \(|1\rangle\), 
\(|2\rangle\) and \(|3\rangle\). Atoms in two of these states, \(|1\rangle\) and \(|2\rangle\), interact reasonably 
strongly.
Components 1 and 2 can be different internal (e.g.\ hyperfine) 
states of an atom, or different 
atomic species (such as $^6$Li and $^{40}$K). Components 2 and 3 must be different internal states of the same 
atom since they have to be coupled by an electromagnetic field (for instance RF-field, or laser fields in a Raman 
configuration).
Even if only pairing, not superfluidity, is necessary for the cooling, we assume here for convenience that the atoms in 
states \(|1\rangle\) and \(|2\rangle\) 
form a superfluid described by the standard BCS-theory with the BCS-Hamiltonian $H_S$. 
Atoms in the third state \(|3\rangle\) are in the normal state, with the corresponding Hamiltonian $H_N$. 
Chemical potentials $\mu_S$ and $\mu_N$ will be added to $H_S$ and $H_N$, and 
incorporate also possible Hartree energies due to weak interactions 
between components $|3\rangle$ and $|1\rangle$ or $|2\rangle$, for definitions see \cite{Torma2000a, Bruun2001a}.
An electromagnetic field couples the 
states \(|2\rangle\) and \(|3\rangle\). This atom-field coupling is described in the rotating wave approximation, and the
total Hamiltonian becomes
\begin{equation}
  H = H_N + H_S + \frac{\delta}{2}(N_S - N_N) + H_{T}.
\end{equation}
The atom-field coupling (tunneling) part of the Hamiltonian is \(H_{T}=\sum_k \Omega \, \hat{c}_{k 2}^\dagger \hat{c}_{k 3} + 
\Omega^* \, \hat{c}_{k 3}^\dagger \hat{c}_{k 2} \equiv \sum_k \Omega \, \hat{c}_{k S}^\dagger \hat{c}_{k N} + 
\Omega^* \, \hat{c}_{k N}^\dagger \hat{c}_{k S}\) ($\hbar = 1$). 
We treat $H_T$ as a perturbation and 
calculate the rate of change of the number of atoms in the superfluid state, 
$\langle \dot{N}_S\rangle=-\langle \dot{N}_N\rangle = \frac{i}{\hbar} \langle [H,N_S] \rangle \equiv \langle \dot{N}\rangle$, 
in the lowest order, i.e.\ using linear response theory \cite{Mahan}. From this we obtain the number of transferred atoms
at time $\tau$ as $\langle \Delta N(\tau)\rangle = \int_0^{\tau}\langle \dot{N}(t') 
\rangle dt'$. The delta-functions which enforce energy conservation in 
$\langle \dot{N} \rangle$ are approximated by (Cauchy-)Lorentz distributions 
of the width $\Gamma$, ${\cal L}(x) = \frac{1}{\pi}\frac{\Gamma}{x^2+{\Gamma}^2}$. 
We assume that the intrinsic linewidth
of the transition is small and the total linewidth is essentially given by the pulse Rabi frequency, 
or inverse pulse length, i.e.\ $\Gamma \sim \Omega$. (The results are not sensitive to the lineshape 
as shown in the following by comparing results for Gaussian and Lorentz distributions). We set $\Gamma = \Omega$, and 
$\Omega \tau = \pi$ corresponding to a $\pi$-pulse which maximizes the transfer. This leads to   
\begin{equation}
\begin{split}
\label{DeltaN_2}
  & \langle \Delta N (\tau = \pi/\Omega)\rangle = 2\pi \sum _k \\ 
  & \frac{\Omega^2}{(E_k - \xi _k^S + \delta)^2 + \Omega^2}u_k^2 \,
  \big(n_{F,N}(\xi _k^N) - n_{F,S}(E_k)\big) + \\  
  & \frac{\Omega^2}{(E_k + \xi _k^S - \delta)^2 + \Omega^2}v_k^2 \,
  \big(n_{F,N}(\xi _k^N) - n_{F,S}(-E_k) \big).
\end{split}
\end{equation}
Here \(\xi _k^{N/S} = \frac{\hbar^2 k^2}{2m} - \mu _{N/S}\) and 
\(E_k = \sqrt{(\xi _k^S)^2 + \Delta^2}\). The number of thermal quasiparticles is given by the Fermi
distribution $1/(e^{E/k_B T} + 1)$; in the paired state $E = E_k$ with the minimum value $\Delta$ whereas 
in the normal state $E = \xi_k^N$ and there is no gap, the distributions in the paired and normal states are 
therefore different.
In the following, 
unless stated otherwise, we assume $\mu_N = \mu_S$, $T_N = T_S$ and set $\hbar=\epsilon _F=k_B=1$.
For clarity, we keep $\Gamma$ and $\Omega$ explicit although in the numerics
we set $\Gamma = \Omega$. 
The Eq.(\ref{DeltaN_2}) shows explicitly the processes illustrated in 
Fig.1.
The term proportional to $u_k$ corresponds to 
the single quasiparticle transfer in Fig.~1b,
and the term proportional to $v_k$ to the pair breaking/formation process in Fig.~1a. The differences in the quasiparticle
distributions, given by $n_{F,N}(\xi_k^N)$ and $n_{F,S}(E_k)$, determine the direction of the net particle transfer. 
For pairing gap $\Delta=0$, 
there is no net particle transfer and no net cooling. 

We also derive the heat flux out of the normal state (i.e.\ the cooling power) 
from $\langle \dot{Q}_N \rangle = \langle \dot{N} H_N \rangle$ 
in the same way as $\langle \dot{N} \rangle$ above.
Since $H_N$ measures energy from the chemical potential $\mu_N$, the energy of removed 
(inserted) hot quasiparticles enters with the same sign as inserted (removed) particles that fill (create) 
holes below the Fermi level. Therefore $\langle \dot{Q}_N \rangle$ directly tells about the amount of cooling. 
In the case of solid state SN junctions, the heat flux $\dot{Q}$ is 
calculated in the same way but the functional dependence on
the relevant parameters differs due to the momentum conservation in our case. 
We obtain
\begin{equation}
\label{heatflux}
\begin{split}
  & \langle \Dot{Q}_N \rangle = 2\pi\Omega^2 \sum_k \xi _k^N \bigg\{ \\ 
  & u_k^2 \, \mathcal{L} \left( \xi _k^S - E_k - \delta \right) \left[ n_{F,N}(\xi _k^N) - n_{F,S}(E_k) \right] + \\
  & v_k^2 \, \mathcal{L} \left( \xi _k^S + E_k - \delta \right) \left[ n_{F,N}(\xi _k^N)-  n_{F,S}(-E_k) \right] \bigg\} .
\end{split}
\end{equation}

To determine the final temperature of the normal state gas after the cooling pulse,
we assume thermalization into a Fermi
distribution after the pulse (for instance due to weak p-wave interactions),
or that the distribution produced by the cooling pulse is close to a Fermi 
distribution. We then obtain the final  
temperature \(\tilde{T}\) and the chemical potential \(\tilde{\mu}\) from 
\(\langle N_N \rangle + \langle \Delta N \rangle = \sum_k \tilde{n}_F(\tilde{\xi} _k)\) and 
\(\langle E_N \rangle + \langle \Delta E \rangle = \sum_k \frac{\hbar^2 k^2}{2m} \tilde{n}_F(\tilde{\xi} _k)\),
where \(E_N = \sum_k \frac{\hbar^2 k^2}{2m} n_{F,N}(\xi _k^N)\), $\tilde{n}_F$ is the Fermi function
with $\tilde{T}$ and $\tilde{\mu}$, and $\langle \Delta E \rangle$ is the energy added by transferring the $\langle \Delta 
N \rangle$ particles. 

In Fig.2 is shown the temperature after the cooling pulse, 
as a function of the detuning, for four different
linewidths $\Gamma$. The optimal detuning $\delta$ is somewhat below the pairing gap energy (in general we find
$\delta \sim \Delta/2$), and the optimal linewidth 
$\Gamma (\sim \Omega)$ is of the same order of magnitude. 
We have tested also temperatures much smaller than the gap: then there is nearly no transfer of particles below the gap energy due to the 
lack of thermal quasiparticles, moreover, the cooling effect is minute and only heating is caused for larger detunings. 
In other words, the pairing gap and the temperature should be of the same order of magnitude, $\Delta \sim T$. In the limits 
$T>>\Delta$ and $T<<\Delta$ the desired effect vanishes. 
\begin{figure}
\label{graph_T20}
\centering
\includegraphics[width = 0.35\textwidth]{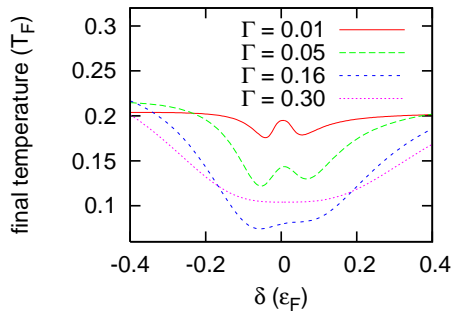}
\caption{(Color online) Temperature of the normal state after the cooling pulse as a function of detuning, for different linewidths
\(\Gamma = 0.01\),  $0.05$, $0.16$ and \(0.30 \epsilon _F\).
Before the pulse, the temperature was \(0.2T_F\) and the pairing gap is \(0.2\epsilon _F\). 
The linewidth \(0.16 \pm 0.01 \epsilon _F\) gives the maximal cooling.}
\end{figure}

Note that for larger linewidths, the cooling works also for zero detuning $\delta$. Since 
$\delta$ plays the same role as voltage applied over an SN contact, this corresponds to cooling in presence of no external 
voltage. Such a system has been discussed recently in an interesting way in \cite{Pekola2007a} 
where cooling by an SN junction is 
driven solely based on thermal noise from a resistor -- the system is viewed as a Brownian refrigerator which conveys heat 
unidirectionally in response to external noise, in analogy to Brownian motors and thermal ratchets or to the concept of 
Maxwell's demon. The effect of the thermal noise is to relax the energy conservation requirement in the tunneling process
-- in our case, this is done by the finite linewidth $\Gamma$ 
of the coupling field. 
We have derived an analytical estimate for the heat flux $\langle \dot{Q}_{N} \rangle$ defined in Eq.(\ref{heatflux}), 
using the following approximations: the Fermi distributions are approximated by their Boltzmann-like tails, the Lorentz
distribution is replaced by a Gaussian of the same width, and the temperature and the linewidth are assumed to be 
smaller than the energy gap, also the detuning should be in the limits \(|\delta| \lesssim \Delta \). The result becomes 
\begin{equation}
\label{analytical}
\begin{split}
  & \frac{\dot{Q}_N}{N_N} = \frac{3\pi\hbar \Omega^2 \sqrt{\mu}}{2\epsilon _F^{3/2}} e^{-\frac{\Delta}{T}} 
  \bigg\{ 
  e^{\left( \frac{\Gamma^2}{2T^2} + \frac{|\delta|}{T} \right)} \times \\ 
  & \left[ \frac{\Gamma^4}{\Delta T^2} + 2 \Delta - 3 |\delta| + \frac{\delta^2}{\Delta} - \frac{3 \Gamma^2}{T}
     + \frac{2 \Gamma^2 |\delta|}{\Delta T} + \frac{\Gamma^2}{\Delta}\right]  \\
  & - e^{-\frac{\Delta}{T_S}+\frac{\Delta}{T}}\left[ 2 \Delta - 3 |\delta| + \frac{\delta^2}{\Delta} + \frac{\Gamma^2}{\Delta}\right]
  \bigg\}
\end{split}
\end{equation}
Here $T\equiv T_N$.
The result can be compared to the one in \cite{Pekola2007a} by substituting \(\Gamma^2 \rightarrow \sigma = 2T_R E_c\) and 
\(\delta \rightarrow E_c\). In \cite{Pekola2007a} the variance $\sigma$ for the 
voltage arises from the temperature of the environment $T_R$, and contains the charging energy $E_c$ due to the junction
and other capacitances. All the exponential factors are the
same in our result and \cite{Pekola2007a}, however, the other terms differ, originating from the momentum conservation
in our case. The analytical estimate (\ref{analytical}) is compared to numerical 
results in Fig.3 by plotting the numerically calculated heat fluxes
for a Gaussian and a  Lorentz distribution, and the approximated result (\ref{analytical}) 
for a Gaussian distribution. 
Fig.3 shows that the analytical estimate 
is reasonably good around the optimal detuning for cooling, however,
one should use it with care for $\delta \simeq 0$.
From Eq.(\ref{analytical}) one can also estimate the
optimal value for the linewidth, in two ways: one can either require a) the argument of the exponential to be zero,
or b) those terms in the prefactor that are negative to be minimized. 
This gives $\Gamma = \sqrt{xT(\Delta + \delta)}$ where 
$x=2$ for case a) and $x=1$ for case b). Notably, this is the same condition as obtained in \cite{Pekola2007a} 
for the optimal amount of ``energy uncertainty'' (variance $\sigma \propto T_R$) in the cooling process. 
For the parameters of Fig.2 the optimal linewidth given by these estimates is between $0.2 \epsilon_F$ (a) and $0.14 \epsilon_F$ (b)
which is in excellent agreement with the numerically found optimal value $0.16 \pm 0.01 \epsilon_F$. 
\begin{figure}
\label{HeatFlux}
\centering
\includegraphics[width = 0.35\textwidth]{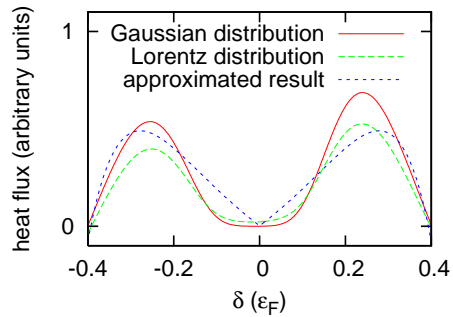}
\caption{(Color online) Energy relative to chemical potential transferred out of the normal state (heat flux), 
$\langle \dot{Q}_{N} \rangle$, 
as a function of the detuning $\delta$. The curves are numerical results for  
Gaussian and Lorentz distributions, and approximate analytical result for a Gaussian distribution.
The gap \(\Delta = 0.4\epsilon _F\), \(T=0.1T_F\) and the linewidth 
\(\Gamma = 0.03\epsilon _F\).}
\end{figure}

Similarly, the heat flux out of the paired state becomes
\begin{equation}
\begin{split}
  & \frac{\dot{Q}_S}{N_N} = \frac{3\pi \Omega^2\sqrt{\mu}}{2} 
  \bigg[ 
  (2\Delta-|\delta|) \times \\
  & \left(
    e^{-\frac{\Delta}{T_S}} - e^{-\frac{\Delta}{T_N}
    +\frac{|\delta|}{T_N} + \frac{\Gamma^2}{2T_N^2}} 
  \right)
  +\frac{\Gamma^2}{T_N} e^{-\frac{\Delta}{T_N} + \frac{|\delta|}{T_N} + \frac{\Gamma^2}{2T_N^2}}
  \bigg] ,
\end{split}
\end{equation}
where one can approximate the last term to be zero when $\Gamma^2/T_N << \Delta$.  
One can see from here that, for suitable values of $T_N<T_S$, it is possible to cool the paired state as well. However,
if $T_N\geq T_S$ cooling of the paired state is not possible because the factor $\frac{|\delta|}{T_N}+\frac{\Gamma^2}{2T_N^2}$ is 
always positive. This is in contrast to the case of \cite{Pekola2007a} where the superconductor can be cooled 
(and the normal metal heated) also for $T_N = T_S$ in the case that the external resistor connected to the SN junction 
has a temperature $T_R$ lower than $T_N$. This illuminates the difference between the external reservoir (resistor) and the 
electromagnetic field: although thermal fluctuations of the reservoir in \cite{Pekola2007a} and 
the linewidth of the field in our case have the same role in 
relaxing the energy conservation in tunneling, the reservoir can also increase its entropy in the process (if $T_R < T_N$) whereas
the electromagnetic field does not have a role in the trade of entropy. In our case the entropy decreased by cooling the normal state
simply increases the entropy of the paired state, and vice versa. E.g. if normal state is cooled from 
0.2 $T_F$ to 0.1 $T_F$ the paired state heats from 0.2 $T_F$ to 0.29 $T_F$. 
To continue the cooling efficiently, the hot quasiparticles produced by the cooling pulse could be selectively 
removed away from the paired state, to a state that is not confined in the trap/lattice.

The proposed cooling method is not sensitive to having the chemical potentials equal, $\mu_S = \mu_N$, nor to homogeneity or
completeness of the gap. This is demonstrated in Fig.4 where we show the heat flux for a non-balanced case 
$\mu_N \neq \mu_S$: cooling works even better, it only becomes asymmetric with respect to the detuning $\delta$.
In practice,
the paired state may well be
different from the homogeneous space BCS-state used here, for instance the (pseudo)gap may be non-complete (i.e.\ it may have some 
spectral weight
at zero energy as well), x-dependent, k-dependent, etc. In Fig.4 we show a linear combination of results for
different gaps, simulating nodes and inhomogeneities in the gap in the spirit of local density approximation. 
The cooling is not dramatically affected, it is simply reduced in proportion with 
the inhomogeneity of the gap. The scheme also works for molecules but is not optimal because the 
pairing energy in that case easily becomes much larger than the temperature. In general, the cooling is based on coupling 
systems with different spectral densities which can be caused also by other effects than pairing, e.g.\ by Landau quantization 
\cite{Giazotto2007a}. 
\begin{figure}
\label{NonBalanced}
\centering
\includegraphics[width = 0.35\textwidth]{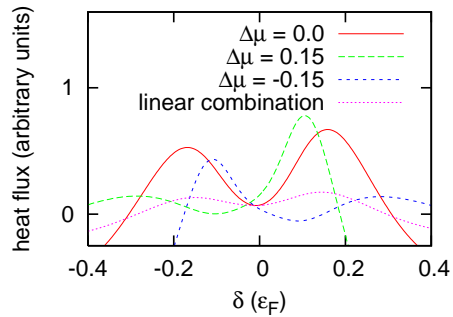}
\caption{(Color online) The heat flux $\langle \dot{Q}_N \rangle$ out of the normal state. The curves are for various  
chemical potential differences $\Delta \mu = \mu_S - \mu_N$ between the paired 
and the normal state. 
The gap $\Delta = 0.3\epsilon _F$, $T=0.1T_F$ and the linewidth $0.05\epsilon _F$. 
A linear combination of heat fluxes with different energy gaps, 
\(\dot{Q}_N = 0.1\dot{Q}_N(\Delta=0) + 0.3\dot{Q}_N(\Delta = 0.15) + 0.6\dot{Q}_N(\Delta=0.3)\), is also shown.}
\end{figure}
    
The setting described here can be used for thermometry as well, as is also done in the solid state context \cite{Giazotto2006a}. 
The number of transferred particles becomes exponentially sensitive to temperature; calculating it in the same way 
as above gives the analytical estimate 
\(\frac{\langle \Delta N_N \rangle}{N_N} \simeq \frac{3\pi^2\Omega\sqrt{\mu}}{2} \, \text{sgn}(\delta)  
   \bigg[ e^{-\frac{\Delta}{T_N} + \frac{|\delta|}{T_N}
  + \frac{\Gamma^2}{2T_N^2}} \times \left(2 - \frac{|\delta|}{\Delta} - \frac{\Gamma^2}{\Delta T_N^2} \right) 
   + e^{-\frac{\Delta}{T_S}} \left(2 - \frac{|\delta|}{\Delta} \right) \bigg] .\)

In summary, we have proposed cooling of a normal state Fermi gas, based on paired state -- normal state tunneling created by
electromagnetic fields. We have determined the optimal parameters for cooling and shown that temperature drops of the order 
of the pairing gap can be achieved. The cooling of a normal Fermi gas may find 
several applications. A promising one is to produce a very cold single-component Fermi gas, 
and subsequently turn on p-wave interactions by Feshbach resonance techniques. Especially in
optical lattices, studies of p-wave pairing are of great interest due to the richness of the phase diagram. 
Our work also opens up a new connection and an opportunity for synergy between 
ultracold gases and solid state nanostructures research. In general, cooling-techniques are related to manipulation 
of the quasiparticle distributions of a quantum system,
which is essential in the study of non-equilibrium effects. 

{\it Acknowledgements} We thank T. Esslinger for useful discussions. This work was supported in part 
by the National Graduate 
School in Materials Physics, QUDEDIS, Academy of Finland (project numbers 106299, 213362, 115020) and National Science
Foundation under Grant No. PHY05-51164, and conducted
as part of a EURYI scheme award. See www.esf.org/euryi. J.K.\ acknowledges the support of the Department of Energy, Office of Basic
Energy Sciences via the Chemical Sciences, Geosciences, and Biosciences Division.

\bibliographystyle{apsrev}
\bibliography{paperi.bib}

\begin{thebibliography}{18}
\expandafter\ifx\csname natexlab\endcsname\relax\def\natexlab#1{#1}\fi
\expandafter\ifx\csname bibnamefont\endcsname\relax
  \def\bibnamefont#1{#1}\fi
\expandafter\ifx\csname bibfnamefont\endcsname\relax
  \def\bibfnamefont#1{#1}\fi
\expandafter\ifx\csname citenamefont\endcsname\relax
  \def\citenamefont#1{#1}\fi
\expandafter\ifx\csname url\endcsname\relax
  \def\url#1{\texttt{#1}}\fi
\expandafter\ifx\csname urlprefix\endcsname\relax\def\urlprefix{URL }\fi
\providecommand{\bibinfo}[2]{#2}
\providecommand{\eprint}[2][]{\url{#2}}

\bibitem[{\citenamefont{Anderson et~al.}(1995)}]{Anderson1995a}
\bibinfo{author}{\bibfnamefont{M.~H.} \bibnamefont{Anderson}}
  \bibnamefont{et~al.}, \bibinfo{journal}{Science}
  \textbf{\bibinfo{volume}{269}}, \bibinfo{pages}{198} (\bibinfo{year}{1995}).

\bibitem[{\citenamefont{Davis et~al.}(1995)}]{Davis1995a}
\bibinfo{author}{\bibfnamefont{K.~B.} \bibnamefont{Davis}}
  \bibnamefont{et~al.}, \bibinfo{journal}{Phys. Rev. Lett.}
  \textbf{\bibinfo{volume}{75}}, \bibinfo{pages}{3969} (\bibinfo{year}{1995}).

\bibitem[{\citenamefont{Greiner et~al.}(2002)}]{Greiner2002a}
\bibinfo{author}{\bibfnamefont{M.}~\bibnamefont{Greiner}} \bibnamefont{et~al.},
  \bibinfo{journal}{Nature} \textbf{\bibinfo{volume}{415}}, \bibinfo{pages}{39}
  (\bibinfo{year}{2002}).

\bibitem[{\citenamefont{Jochim et~al.}(2003)}]{Jochim2003b}
\bibinfo{author}{\bibfnamefont{S.}~\bibnamefont{Jochim}} \bibnamefont{et~al.},
  \bibinfo{journal}{Science} \textbf{\bibinfo{volume}{302}},
  \bibinfo{pages}{2101} (\bibinfo{year}{2003}).

\bibitem[{\citenamefont{Regal et~al.}(2003)}]{Greiner2003a}
\bibinfo{author}{\bibfnamefont{C.}~\bibnamefont{Regal}} \bibnamefont{et~al.},
  \bibinfo{journal}{Nature} \textbf{\bibinfo{volume}{426}},
  \bibinfo{pages}{537} (\bibinfo{year}{2003}).

\bibitem[{\citenamefont{Regal et~al.}(2004)}]{Regal2004a}
\bibinfo{author}{\bibfnamefont{C.}~\bibnamefont{Regal}} \bibnamefont{et~al.},
  \bibinfo{journal}{Phys. Rev. Lett.} \textbf{\bibinfo{volume}{92}},
  \bibinfo{pages}{040403} (\bibinfo{year}{2004}).

\bibitem[{\citenamefont{Zwierlein et~al.}(2004)}]{Zwierlein2004a}
\bibinfo{author}{\bibfnamefont{M.~W.} \bibnamefont{Zwierlein}}
  \bibnamefont{et~al.}, \bibinfo{journal}{Phys. Rev. Lett.}
  \textbf{\bibinfo{volume}{92}}, \bibinfo{pages}{120403}
  (\bibinfo{year}{2004}).

\bibitem[{\citenamefont{Bartenstein et~al.}(2004)}]{Bartenstein2004a}
\bibinfo{author}{\bibfnamefont{M.}~\bibnamefont{Bartenstein}}
  \bibnamefont{et~al.}, \bibinfo{journal}{Phys. Rev. Lett.}
  \textbf{\bibinfo{volume}{92}}, \bibinfo{pages}{120401}
  (\bibinfo{year}{2004}).

\bibitem[{\citenamefont{Kinast et~al.}(2004)}]{Kinast2004a}
\bibinfo{author}{\bibfnamefont{J.}~\bibnamefont{Kinast}} \bibnamefont{et~al.},
  \bibinfo{journal}{Phys. Rev. Lett.} \textbf{\bibinfo{volume}{92}},
  \bibinfo{pages}{150402} (\bibinfo{year}{2004}).

\bibitem[{\citenamefont{Chin et~al.}(2004)}]{Chin2004a}
\bibinfo{author}{\bibfnamefont{C.}~\bibnamefont{Chin}} \bibnamefont{et~al.},
  \bibinfo{journal}{Science} \textbf{\bibinfo{volume}{305}},
  \bibinfo{pages}{1128} (\bibinfo{year}{2004}).

\bibitem[{\citenamefont{Zwierlein et~al.}(2005)}]{Zwierlein2005a}
\bibinfo{author}{\bibfnamefont{M.~W.} \bibnamefont{Zwierlein}}
  \bibnamefont{et~al.}, \bibinfo{journal}{Nature}
  \textbf{\bibinfo{volume}{435}}, \bibinfo{pages}{1047} (\bibinfo{year}{2005}).

\bibitem[{\citenamefont{Kinnunen et~al.}(2004)}]{Kinnunen2004b}
\bibinfo{author}{\bibfnamefont{J.}~\bibnamefont{Kinnunen}}
  \bibnamefont{et~al.}, \bibinfo{journal}{Science}
  \textbf{\bibinfo{volume}{305}}, \bibinfo{pages}{1131} (\bibinfo{year}{2004}).

\bibitem[{\citenamefont{Giazotto et~al.}(2006)}]{Giazotto2006a}
\bibinfo{author}{\bibfnamefont{F.}~\bibnamefont{Giazotto}}
  \bibnamefont{et~al.}, \bibinfo{journal}{Rev. Mod. Phys.}
  \textbf{\bibinfo{volume}{78}}, \bibinfo{pages}{217} (\bibinfo{year}{2006}).

\bibitem[{\citenamefont{Mahan}(2000)}]{Mahan}
\bibinfo{author}{\bibfnamefont{G.~D.} \bibnamefont{Mahan}},
  \emph{\bibinfo{title}{Many-Particle Physics}} (\bibinfo{publisher}{Kluwer
  Academic/Plenum Publishers}, \bibinfo{address}{New York},
  \bibinfo{year}{2000}).

\bibitem[{\citenamefont{T{\"o}rm{\"a} and Zoller}(2000)}]{Torma2000a}
\bibinfo{author}{\bibfnamefont{P.}~\bibnamefont{T{\"o}rm{\"a}}}
  \bibnamefont{and} \bibinfo{author}{\bibfnamefont{P.}~\bibnamefont{Zoller}},
  \bibinfo{journal}{Phys. Rev. Lett.} \textbf{\bibinfo{volume}{85}},
  \bibinfo{pages}{487} (\bibinfo{year}{2000}).

\bibitem[{\citenamefont{Bruun et~al.}(2001)}]{Bruun2001a}
\bibinfo{author}{\bibfnamefont{G.~M.} \bibnamefont{Bruun}}
  \bibnamefont{et~al.}, \bibinfo{journal}{Phys. Rev. A}
  \textbf{\bibinfo{volume}{64}}, \bibinfo{pages}{033609}
  (\bibinfo{year}{2001}).

\bibitem[{\citenamefont{Pekola and Hekking}(2007)}]{Pekola2007a}
\bibinfo{author}{\bibfnamefont{J.~P.} \bibnamefont{Pekola}} \bibnamefont{and}
  \bibinfo{author}{\bibfnamefont{F.~W.~J.} \bibnamefont{Hekking}}
  (\bibinfo{year}{2007}), \eprint{cond-mat/0702233}.

\bibitem[{\citenamefont{Giazotto et~al.}(2007)}]{Giazotto2007a}
\bibinfo{author}{\bibfnamefont{F.}~\bibnamefont{Giazotto}} \bibnamefont{et~al.}
  (\bibinfo{year}{2007}), \eprint{cond-mat/0703119}.

\end{thebibliography}
\end{document}